\documentclass[aps,prl,twocolumn]{revtex4}
\usepackage[]{graphicx}
\newcommand{\dd}{\mathrm d}

\usepackage{color}

\begin{document}

\title{Better stability with measurement errors}

\author{Aykut Argun}
\affiliation{Department of Physics, Bilkent University, Cankaya, Ankara 06800, Turkey.}

\author{Giovanni Volpe}
\affiliation{Department of Physics, Bilkent University, Cankaya, Ankara 06800, Turkey.}

\date{\today}

\begin{abstract}
Often it is desirable to stabilize a system around an optimal state. This can be effectively accomplished using feedback control, where the system deviation from the desired state is measured in order to determine the magnitude of the restoring force to be applied. Contrary to conventional wisdom, i.e. that a more precise measurement is expected to improve the system stability, here we demonstrate that a certain degree of measurement error can improve the system stability. We exemplify the implications of this finding with numerical examples drawn from various fields, such as the operation of a temperature controller, the confinement of a microscopic particle, the localization of a target by a microswimmer, and the control of a population.
\end{abstract}


\maketitle

The presence of noise has a deleterious effect on many phenomena as it can drive a system away from its  optimal or desired working conditions \cite{golnaraghi2008automatic}. For example, Brownian fluctuations have to be fought by microscopic organisms, e.g. cells and bacteria, in their search for food and mates \cite{berg2004coli}; and environmental fluctuations can alter the equilibrium of an ecosystem and must be taken into account, e.g. in the management of endangered species and of fisheries \cite{botsford1997management,meffe2012ecosystem}. In these situations, feedback control is a powerful technique to stabilize a system, where the system deviation from the desired state is measured in order to determine the magnitude of the restoring force to be applied \cite{franklin1991feedback}. The quality of the feedback control depends on the quality of the system state measurement: in principle, one could expect that a more accurate measurement should lead to a better system stability. However, we will show that, when the restoring force grows more than linearly with the deviation, the system stability improves in the presence of measurement errors. This result permits one to engineer the right conditions to relax the requirements, and therefore the cost, of the measurement procedures.

\begin{figure}[b]
\includegraphics*[width=3.25in]{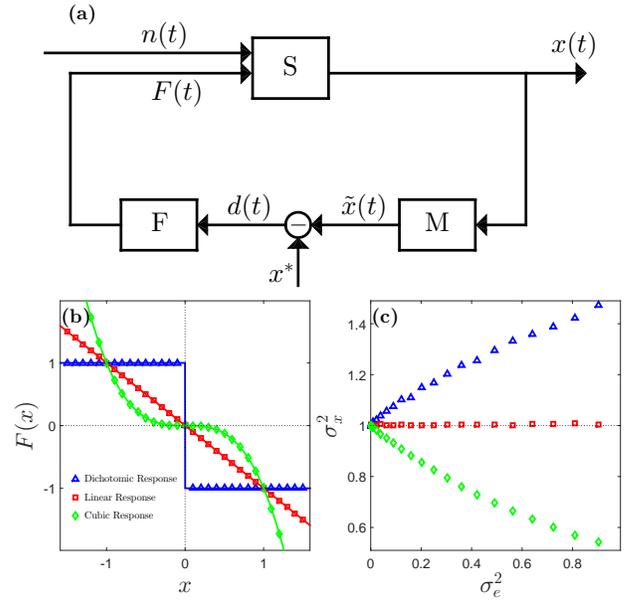}
\caption{(Color online) (a) Schematic view of a noisy system whose state $x(t)$ evolves in time under the influence of a noise $n(t)$. A feedback force $F(t)$ acts on the system to keep it as close as possible to the optimal state $x^*$. $F(t)$ is proportional to the deviation $d(t) = \tilde{x}(t)-x^*$ between the measured system state $\tilde{x}(t) = x(t)+e(t)$ and $x^*$, where $e(t)$ is the measurement error.
(b) Dichotomic (triangles, $\alpha = 0$), linear (squares, $\alpha = 1$), and cubic (diamonds, $\alpha = 3$) feedback forces calcualted according to Eq.~(\ref{eq:Falpha}).
(c) Numerical results for the system variance $\sigma_x^2$ as a function of the measurement error variance $\sigma_e^2$ for $\alpha=0$ (triangles), $1$ (squares), and $3$ (diamonds).
}
\label{fig1}
\end{figure}

As a model system (Fig.~\ref{fig1}(a)), we consider a one-dimensional dynamic system whose state $x(t)$ evolves in time under the influence of some random fluctuations. These fluctuations can be modeled by a noisy driving term $n(t)$, which we will assume to be a Gaussian white noise with zero mean and variance $\sigma_n^2$. In order to keep $x(t)$ as close as possible to its optimal state $x^*$, we introduce a feedback loop consisting of the following steps:
\begin{enumerate}
\item measurement of the current system state $\tilde{x}(t)$;
\item calculation of the system deviation from $x^*$, i.e. $d(t) = \tilde{x}(t)-x^*$;
\item application of a restoring force depending on $d(t)$, i.e. $F(d(t))$.
\end{enumerate}
In general, the measured system state $\tilde{x}(t)$ is different from the real instantaneous system state $x(t)$, i.e. there is a \emph{measurement error}
\begin{equation}\label{eq:e}
e(t) = \tilde{x}(t) - x(t),
\end{equation}
which we will assume to have zero average and variance $\sigma_e^2$, to be stationary, and to fluctuate on a timescale $\tau_e$ significantly shorter than the system oscillations around its equilibrium position. The resulting system dynamics are described by the first-order stochastic differential equation (SDE) \cite{oksendal2003stochastic}
\begin{equation}\label{eq:2}
\frac{d}{dt} x(t) = F(\tilde{x}(t)-x^*) + n(t).
\end{equation}
In order to evaluate the system stability, we will consider the variance of $x$ around $x^*$:
\begin{equation}
\sigma_x^2 = \overline{\left(x(t)-x^*\right)^2},
\end{equation}
where the overline represents a time average. The more stable the system is, the smaller its variance will be \cite{PhysRevLett.86.950,volpe2009thermal}.

The resulting system stability depends on the feedback force. In general, we will consider forces of the form
\begin{equation}\label{eq:Falpha}
F_{\alpha}(d)= -\mathrm{sign}(d) \; C \; \left|\frac{d}{\delta}\right|^\alpha,
\end{equation}
where $\alpha \ge 0$ is a real parameter, $C$ is a positive constant representing the confinement effort, and $\delta$ is a parameter related to the characteristic amplitude of the system state oscillations around its equilibrium. Some examples of feedback forces are illustrated in Fig.~\ref{fig1}(b) and the respective dependence of $\sigma_x^2$ on $\sigma_e^2$ in Fig.~\ref{fig1}(c). When $\alpha = 0$, the feedback force is dichotomic, i.e. it depends only on the sign of $d$ (triangles in Fig.~\ref{fig1}(b)), and $\sigma_x^2$ monotonically increases with $\sigma_e^2$ (triangles in Fig.~\ref{fig1}(c)). Similar results are obtained for $\alpha \le 1$; in particular, for $\alpha = 1$ the feedback force is linear in $d$ (squares in Fig.~\ref{fig1}(b)) and $\sigma_x^2$ increases with $\sigma_e^2$ (squares in Fig.~\ref{fig1}(c)), even though in this case the slope is weaker and, as will be shown below, as $\tau_e \rightarrow 0$, $\sigma_x^2$ becomes independent from $\sigma_e^2$. Finally, the most interesting case is when $\alpha>1$, i.e. when the feedback force grows more than linearly with $d$: when $\sigma_e^2$ increases, $\sigma_x^2$ decreases, as illustrated for $\alpha = 3$ by the diamonds in Figs.~\ref{fig1}(b) and \ref{fig1}(c). Therefore, for $\alpha>0$, we obtain the counterintuitive result that the system stability increases as the quality of the system state measurement decreases.

In order to understand the nature of this result, we will first consider the case when a perfect measurement of the system state is possible, i.e. $\sigma_e^2 = 0$. In this case, $e(t)\equiv 0$ and the SDE describing the system is
\begin{equation}\label{eq:SDEnonoise}
\frac{d}{dt} x(t) = F(x(t)-x^*) + n(t).
\end{equation}
We can now use the fact that $F(x)$ is associated to the potential $U^{(0)}(x) = -\int_{x^*}^x F(y-x^*)dy$ and therefore the probability distribution of the system states is
\begin{equation}\label{eq:p0}
p^{(0)}(x)= \frac{\exp \left\{ -\beta U^{(0)}(x) \right\} }{Z} = \frac{\exp \left\{ \beta \int_{x^*}^{x} F(y-x^*) dy \right\} }{Z},
\end{equation}
where $\beta = 2\sigma_n^{-2}$ is proportional to the inverse temperature and $Z = \int \exp \left\{ \beta \int_{-\infty}^{+\infty} F(y-x^*) dy \right\} dx$ is the partition function, to calculate the variance $\sigma_x^{2,(0)}$ for the process described by Eq.~(\ref{eq:SDEnonoise}) as
\begin{equation}\label{eq:sx0}
\sigma_{x}^{2,(0)} = \int_{-\infty}^{+\infty} \left(x-x^*\right)^2 p^{(0)}(x) dx,
\end{equation}
where the superscripts ``$(0)$" have been added as a reminder that these quantities correspond to a system without measurement noise.

We now consider the case when a measurement error is present, i.e. $\sigma_e^2 \neq 0$. Since we have assumed the correlation time of the measurement error $\tau_e$ to be much smaller than the characteristic time scales of the system, for each system state $x$ we can introduce an effective force that averages the various measurement noises and, thus, depends only on $x$. This permits us to rewrite Eq.~(\ref{eq:2}) in terms of the system state $x$, i.e.,
\begin{equation}\label{eq:SDEnoise}
\frac{d}{dt} x(t) = F_{\rm eff}(x(t)-x^*) + n(t),
\end{equation}
where
\begin{equation}
F_{\rm eff}(x-x^*) = \int_{-\infty}^{\infty} F(x-x^*+e) p_{e}(e) de.
\end{equation}
Following the same  procedure used to derive Eqs.~(\ref{eq:p0}) and (\ref{eq:sx0}), we can then obtain the probability distribution of the system state
\begin{equation}\label{eq:p}
p^{(e)}(x) = \frac{\exp \left\{ \beta \int_{x^*}^{x} F_{\rm eff}(y-x^*) dy \right\} }{Z}
\end{equation}
and its variance	
\begin{equation}\label{eq:sx}
\sigma_{x}^{2,(e)} = \int_{-\infty}^{+\infty} \left(x-x^*\right)^2 p^{(e)}(x) d x,
\end{equation}
where the superscripts ``$(e)$" have been added as a reminder that these quantities depend on the measurement noise characteristics. We note at this point that, if $F_{\rm eff}(x) > F(x)$ for all $x$, $p^{(e)}$ is more compact than $p^{(0)}$, and therefore $\sigma_{x}^{2,(e)} < \sigma_{x}^{2,(0)}$. In order to understand what are the conditions for this to apply, we analyze $F_{\rm eff}(x-x^*+e)$. We start by considering the Taylor expansion of $F_{\rm eff}(x-x^*+e)$ around $x-x^*$, which gives
\begin{widetext}
$$
F_{\rm eff}(x-x^*) =
\int_{-\infty}^{\infty}     \left[
F(x-x^*)
+ e \frac{d F(x-x^*)}{d x}
+ \frac{e^2}{2} \frac{d^2 F(x-x^*)}{d x^2}
+ \mathcal{O}(e^3)
\right] p_e(e) de.
$$
From the previous equation, assuming a small noise level and neglecting terms in the third power of $e$, we obtain
$$
F_{\rm eff}(x-x^*) =
F(x-x^*)
\underbrace{ \int_{-\infty}^{\infty} p_e(e) de }_{=1}
+ \frac{d F(x-x^*)}{d x}
\underbrace{ \int_{-\infty}^{\infty} e p_e(e) de }_{=0}
+ \frac{1}{2} \frac{d^2 F(x-x^*)}{d x^2}
\underbrace{ \int_{-\infty}^{\infty} e^2 p_e(e) de }_{=\sigma_e^2},
$$
\end{widetext}
and, thus,
\begin{equation}\label{eq:Feff}
F_{\rm eff}(x-x^*) = F(x-x^*) + \frac{\sigma_e^2}{2} \frac{d^2 F(x-x^*)}{d x^2}.
\end{equation}
From Eq.~(\ref{eq:Feff}), we can conclude that $F_{\rm eff}(x-x^*) > F(x-x^*)$ only if $\displaystyle \frac{d^2 F(x-x^*)}{d x^2}>0$. In the case of the forces expressed by Eq.~(\ref{eq:Falpha}), Eq.~(\ref{eq:Feff}) becomes
\begin{equation}
\textstyle F_{{\rm eff},\alpha} (x-x^*) = -\mathrm{sign}(x-x^*) \; C \; \left|\frac{x-x^*}{\delta}\right|^\alpha \;
\left[ 1 + \frac{\sigma_e^2}{2}\frac{\alpha (\alpha -1)}{(x-x^*)^2} \right],
\end{equation}
from which follows that a reduction of the system variance in the presence of measurement errors is possible only for $\alpha>1$. This is in agreement with the numerical results presented in Fig.~\ref{fig1}(c).

\begin{figure}[b]
\includegraphics*[width=3.25in]{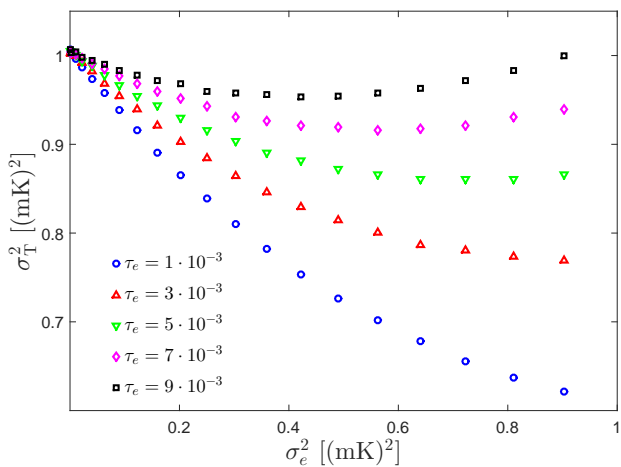}
\caption{(Color online) Decrease of the variance $\sigma_T^2$ of the temperature of a device controlled by a cubic feedback ($\alpha=3$) as a function of the temperature measurement error variance $\sigma_e^2$. As the correlation time of the measurement noise $\tau_e$ increases, the decrease of $\sigma_T^2$ lessens. Each data point is obtained by simulating Eq.~(\ref{eq:TC}) for $20\,000\,{\rm s}$ and with $T^*=300\,{\rm K}$, $K= 1\,{\rm J/K}$, $C_T=2.3 \cdot 10^7\,{\rm W}$, $\Delta T = 1\,{\rm K}$, and $\sigma_n^2 = 10^{-4}\,{\rm K^2}$.}
\label{fig2}
\end{figure}

In order to understand the implications of our result, we now consider some concrete numerical examples where it can find application  \cite{kloeden,volpe2013simulation}. The first example is a temperature controller that must keep a device with heat capacity $K$ at the optimal working temperature $T^*$. The system temperature of the system is $T(t)$. The temperature controller can be realized by using a temperature sensing device, which measures the temperature $\tilde{T}(t) = T(t)+e_T(t)$ with an error $e_T(t)$, and a heating/cooling element with heating/cooling power $C_T$. The resulting equation that describes such a system is
\begin{equation}\label{eq:TC}
K \frac{d T}{d t}= - C_T  \; \mathrm{sign}\left(\tilde{T}(t)-T^*\right) \; \left| \frac{\tilde{T}(t)-T^*}{\Delta T} \right|^{\alpha} + n(t),
\end{equation}
where $\Delta T$ is the characteristic temperature range of the system. Qualitatively, the results for $\alpha=0,1,3$ are the same as the ones presented in Fig.~\ref{fig1}(c); in particular, a decrease of the variance $\sigma_T^2$ of the system is observed for $\alpha=3$. It is interesting, however, to analyze in more detail the role of the noise correlation time $\tau_e$  for $\alpha=3$: as illustrated in Fig.~\ref{fig2}, the decrease of $\sigma_T^2$ as a function of the measurement error becomes smaller as $\tau_e$ increases.

\begin{figure}[b]
\includegraphics*[width=3.25in]{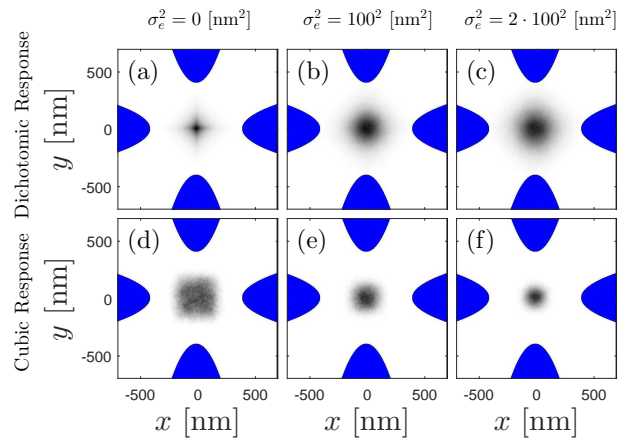}
\caption{(Color online) Histograms of the position of a charged colloidal particle in a optoelectronic tweezers using (a-c) dichotomic and (d-f) cubic feedback. The intensity of the measurement noise increases from left to right. Each histogram is obtained by simulating the motion of a Brownian particle (radius $1\,{\rm \mu m}$, $\gamma = 1.9 \cdot 10^{-8}  \,{\rm N\,s\, m^{-1}}$ ) in a OET using Eq.~(\ref{eq:oet}) for $1000\,{\rm s}$ and $k=5.9 \cdot 10^{-14}\,{\rm N}$ for both dichotomic and cubic feedback. The positional variance is $\sigma_{n,x}^2=\sigma_{n,y}^2=10\,000\,{\rm nm^2}$ in both (a) and (d); $15\,000\,{\rm nm^2} $ in (b) vs. $5\,000\,{\rm nm^2}$ in (e); and $18\,000\,{\rm nm^2}$ in (c) vs. $3\,000\,{\rm nm^2}$ in (f).}
\label{fig3}
\end{figure}

Our central result, i.e. that the presence of measurement errors can improve stability, can also find application in the case of optoelectronic tweezers (OET) \cite{ohta2007optically}. OET are employed to control the motion of microscopic and nanoscopic charged particles by applying an external electric field with the help of electrodes. The intrinsic noise in the particle position emerges as a consequence of Brownian motion, due to the random collisions with the surrounding fluid molecules. OET work by measuring the particle's position, typically using either digital video microcopy \cite{crocker1996methods} or a photodetector \cite{rohrbach2002three}, and by applying a potential difference between the electrodes in order to obtain a restoring electric force. In this case, the motion of the particle in two dimensions can be described by a set of two Langevin equations:
\begin{equation}\label{eq:oet}
\left\{\begin{array}{ccc}
\displaystyle \gamma \frac{dx}{dt} & = & \displaystyle - k \; \mathrm{sign}(\tilde{x}(t)-x^*) \; \left| \frac{\tilde{x}(t)-x^*}{\Delta x} \right|^{\alpha} + W_x(t) \\
\displaystyle \gamma \frac{dy}{dt} & = & \displaystyle - k \; \mathrm{sign}(\tilde{y}(t)-y^*) \; \left| \frac{\tilde{y}(t)-y^*}{\Delta y} \right|^{\alpha} + W_y(t)
\end{array}\right.
\end{equation}
where $[\tilde{x}(t),\tilde{y}(t)] = [x(t)+e_x(t),y(t)+e_y(t)]$ is the measured particle position, $[x(t),y(t)]$ is the particle position, $[e_x(t),e_y(t)]$ is the error in the position measurement, $[x^*,y^*]$ is the desired position, $k$ is the strength of the restoring force, $[\Delta x,\Delta y]$ is the characteristic length scale of the trap, $\gamma$ is the friction coefficient of the particle, $[W_x(t),W)y(t)]$ are uncorrelated white noises with zero mean and variance $2D$, $D=\frac{k_{\rm B}T}{\gamma}$, $T$ is the absolute temperature of the system, and $k_{\rm B}$ is the Boltzmann constant. The results of the corresponding simulations are presented in Fig.~\ref{fig3}. For a dichotomic response ($\alpha=0$, Figs.~\ref{fig3}(a-c)), an increase of the measurement error translates into an increase of the particle variance, as can be seen from the fact that the particle histograms spread over a larger area as $\sigma_e^2$ increases. However, for a cubic response ($\alpha=3$, Figs.~\ref{fig3}(d-f)), the particle confinement improves as the measurement error increases.

\begin{figure}[b]
\includegraphics*[width=3.25in]{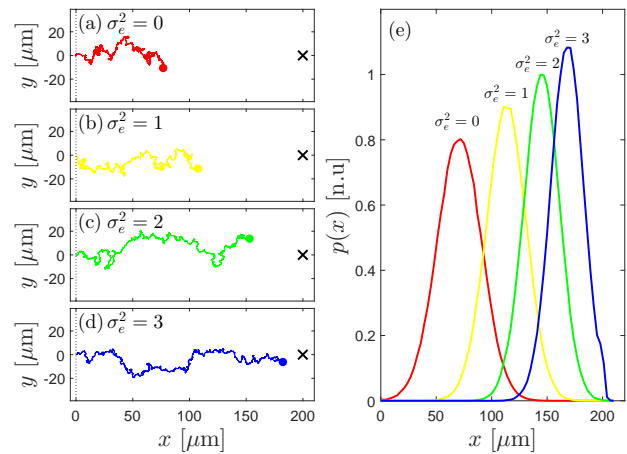}
\caption{(Color online) (a-d) Sample trajectories of microswimmers moving towards a target point (indicated by the cross) as a function of the angular error in the measurement of the propagation direction $\sigma_e^2$. (e) Histograms of the endpoints of the microswimmer trajectory obtained from $350\,000$ simulations. Thanks to the cubic response of the feedback (Eq.~(\ref{eq:micro:fb})), the target is approached more efficiently when more measurement noise is present. The trajectories of the (spherical) microswimmers are simulated for $150\,{\rm s}$ using Eq.~(\ref{eq:micro}) with parameters: $D = k_{\rm B}T/\gamma$, $D_{\rm r} = 3D/(2R)^2$, $\gamma = 6\pi\eta R$, $R = 0.5\,{\rm \mu m}$, $\eta = 0.001\,{\rm Pa\,s}$, $T=300\,{\rm K}$, $k = 0.1\,{\rm s^{-1}}$, and $v = 20\,{\rm \mu \, s^{-1}}$. See also the supplementary video.}
\label{fig4}
\end{figure}

In yet another field, biological and artificial microswimmers are attracting a lot of attention from the biological and physical communities alike as possible candidates for the localization, pick-up, and delivery of microscopic cargoes in microscopic environments \cite{ebbens2010pursuit,bechinger2016active}. In order to perform such tasks, a crucial step is for the microswimmers to be able to reach a certain target using their self-propulsion. A critical problem arises because rotational diffusion prevents a microswimmer from keeping a straight trajectory and forces it to reassess its orientation periodically \cite{berg2004coli}; several strategies have been developed to overcome this problem, including swim-and-tumble chemotaxis \cite{berg1972chemotaxis} and, recently, the use of delayed sensorial feedback \cite{PhysRevX.6.011008}. Here, we consider a microswimmer aiming to reach a target at position $[x_{\rm T},y_{\rm T}]$. The microswimmer is at position $[x(t),y(t)]$ at time $t$ and propels itself with a constant speed $v$ in the direction of its orientation indicated by the angle $\varphi(t)$ \cite{volpe2014simulation}. In order to adjust its orientation towards the target, the microswimmer measures its instantaneous orientation $\tilde{\varphi}(t) = \varphi + e_\varphi$ with an error $e_{\varphi}(t)$ and applies on itself a torque that results in an angular rotation given by 
\begin{equation}\label{eq:micro:fb}
\tau(t) = -k \left( \tilde{\varphi}(t) - \varphi^* \right)^3,
\end{equation}
which is a cubic feedback. The resulting motion of the microswimmer can then be described by the set of SDEs \cite{volpe2014simulation}:
\begin{equation}\label{eq:micro}
\left\{\begin{array}{ccc}
\frac{dx}{dt} & = & v \cos(\varphi (t) )+ \sqrt{2D} W_x(t) \\[6pt]
\frac{dy}{dt} & = & v \sin(\varphi (t) )+ \sqrt{2D} W_y(t) \\[6pt]
\frac{d\varphi (t)}{dt} & = & \tau(\tilde{\varphi}(t),x(t),y(t)) + \sqrt{2D_{\rm r}}W_\varphi (t)
\end{array}\right.
\end{equation}
where $W_x$, $W_y$ and $W_\varphi$ are white noises with zero mean and unitary variance, $D$ is the diffusion coefficient of the microswimmer, and $D_{\rm r}$ is its rotational diffusion coefficient. We examined how fast this swimmer can reach its target depending on measurement errors. As can be seen in Fig.~\ref{fig4}, thanks to the cubic response of the feedback, the microswimmers reaches its target faster when the measurement noise level is higher. 

\begin{figure}[t]
\includegraphics*[width=3.25in]{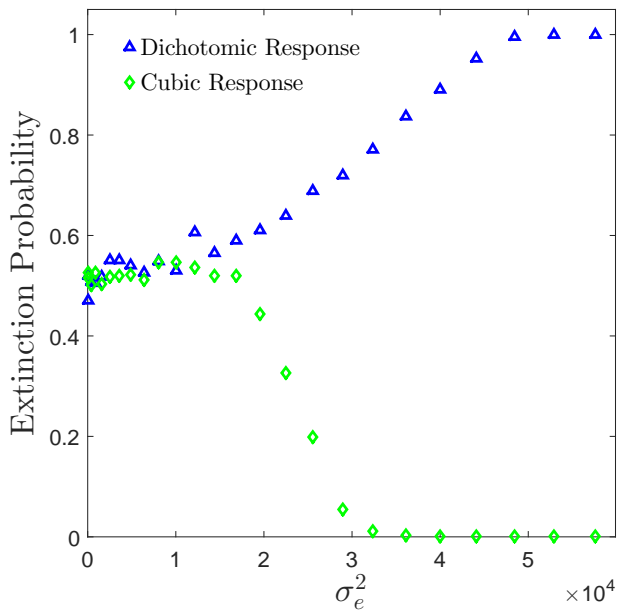}
\caption{(Color online) The extinction rates of fish populations which are controlled with cubic feedback(diamonds) and dichotomic feedback(triangles). The results are obtained from numerical simulations of the equations () and () and extinction probability calculated using 3500 sample runnings over 10 years. Simulations parameters: C (1000) , $R=99,9 (year^{-1})$, $W_R=11.1$, dichotomic response stiffness ($k=3.3342 *10^4$) and cubic feedback stiffness ($k=2 *10^-3$). These values are set in order to meet 50 probability for both cases in the absense of measurement noise.    } 
\label{fig5}
\end{figure}

Finally, we will consider the stabilization of a fishery in order to optimize production. In first approximation, it is crucial to stabilize the population around a level that provides the fastest reproduction. If the resulting population dynamics obey the logistic equation, i.e.
\begin{equation}
\frac{{\dd}x}{{\dd} t}= R x \left(1-{x \over C}\right),
\end{equation}
where $x$ is the population size, $C$ is the carrying capacity and $R$ is the growth rate, production can be optimized by adjusting the fishing rate so that the actual population is equal to $C/2$, which corresponds to the highest population growth rate. We can now consider a more realistic situation where the growth rate is noisy, i.e.
\begin{equation}
R(t)= R_0 + \sigma_R W(t),
\end{equation}
where $R_0= \left<R(t)\right>$, $\sigma_R^2 = \left<(R(t) - R_0)^2\right>$, and $W(t)$ is a white noise. Since now the population tends to deviate from the ideal size, a feedback control should be applied to the fishing rate in order to restore the population back to $C/2$ and, even more importantly, to prevent extinction. The simplest strategy is to apply a dichotomic feedback such that the resulting population dynamic is described by
\begin{equation}
\frac{{\dd}x}{{\dd} t}= R(t)x \left(1-{x \over C}\right) \underbrace{- R(t) \frac{C}{4}}_{\rm fishing} - k \, {\rm sign}\left( \tilde{x}(t)-C/2 \right),
\end{equation}
where $\tilde{x}(t) = x(t)+e(t)$ is the measured population size and $e(t)$ is the error in the assessment of the fish population. If we apply such strategy to an ensemble of fisheries, we obtain that the extinction probability grows to certainty as the measurement error in the assessment of the population grows, as shown by the triangles in Fig.~\ref{fig5}. We can now try and apply a cubic feedback control, so that the resulting population dynamics is described by 
\begin{equation}
\frac{{\dd}x}{{\dd} t}= R(t)x \left(1-{x \over C}\right) \underbrace{- R(t) \frac{C}{4}}_{\rm fishing} - k \, \left( \tilde{x}(t)-C/2 \right)^3.
\end{equation}
In this case, as the measurement error increases the extinction probability goes down to zero, as shown by the squares in Fig.~\ref{fig5}.

In conclusion, we have shown that the presence of noise in the measurement of a system status is not necessarily deleterious and can, in fact, improve system stability depending on the functional form of the feedback response. As a consequence, an addition of noise can effectively reduce the system variance and, therefore, enhance stability.

\begin{acknowledgments}
GV has been partially financially supported by Marie Curie Career Integration Grant (MC-CIG) under Grant PCIG11 GA-2012-321726 and a Distinguished Young Scientist award of the Turkish Academy of Sciences (T\"UBA).
\end{acknowledgments}


\begin{thebibliography}{18}
\expandafter\ifx\csname natexlab\endcsname\relax\def\natexlab#1{#1}\fi
\expandafter\ifx\csname bibnamefont\endcsname\relax
  \def\bibnamefont#1{#1}\fi
\expandafter\ifx\csname bibfnamefont\endcsname\relax
  \def\bibfnamefont#1{#1}\fi
\expandafter\ifx\csname citenamefont\endcsname\relax
  \def\citenamefont#1{#1}\fi
\expandafter\ifx\csname url\endcsname\relax
  \def\url#1{\texttt{#1}}\fi
\expandafter\ifx\csname urlprefix\endcsname\relax\def\urlprefix{URL }\fi
\providecommand{\bibinfo}[2]{#2}
\providecommand{\eprint}[2][]{\url{#2}}

\bibitem[{\citenamefont{Kuo and Golnaraghi}(2008)}]{golnaraghi2008automatic}
\bibinfo{author}{\bibfnamefont{B.~C.} \bibnamefont{Kuo}} \bibnamefont{and}
  \bibinfo{author}{\bibfnamefont{F.}~\bibnamefont{Golnaraghi}},
  \emph{\bibinfo{title}{Automatic control systems}} (\bibinfo{publisher}{John
  Wiley \& Sons, Inc.}, \bibinfo{address}{New York, NY}, \bibinfo{year}{2008}),
  \bibinfo{edition}{8th} ed.

\bibitem[{\citenamefont{Berg}(2004)}]{berg2004coli}
\bibinfo{author}{\bibfnamefont{H.~C.} \bibnamefont{Berg}},
  \emph{\bibinfo{title}{E. coli in Motion}} (\bibinfo{publisher}{Springer
  Verlag}, \bibinfo{address}{Heidelberg, Germany}, \bibinfo{year}{2004}).

\bibitem[{\citenamefont{Botsford et~al.}(1997)\citenamefont{Botsford, Castilla,
  and Peterson}}]{botsford1997management}
\bibinfo{author}{\bibfnamefont{L.~W.} \bibnamefont{Botsford}},
  \bibinfo{author}{\bibfnamefont{J.~C.} \bibnamefont{Castilla}},
  \bibnamefont{and} \bibinfo{author}{\bibfnamefont{C.~H.}
  \bibnamefont{Peterson}}, \bibinfo{journal}{Science}
  \textbf{\bibinfo{volume}{277}}, \bibinfo{pages}{509} (\bibinfo{year}{1997}).

\bibitem[{\citenamefont{Meffe et~al.}(2012)\citenamefont{Meffe, Nielsen,
  Knight, and Schenborn}}]{meffe2012ecosystem}
\bibinfo{author}{\bibfnamefont{G.}~\bibnamefont{Meffe}},
  \bibinfo{author}{\bibfnamefont{L.}~\bibnamefont{Nielsen}},
  \bibinfo{author}{\bibfnamefont{R.~L.} \bibnamefont{Knight}},
  \bibnamefont{and}
  \bibinfo{author}{\bibfnamefont{D.}~\bibnamefont{Schenborn}},
  \emph{\bibinfo{title}{Ecosystem management: Adaptive, community-based
  conservation}} (\bibinfo{publisher}{Island Press},
  \bibinfo{address}{Washington, DC}, \bibinfo{year}{2012}).

\bibitem[{\citenamefont{Franklin et~al.}(1991)\citenamefont{Franklin, Powell,
  and Emami-Naeini}}]{franklin1991feedback}
\bibinfo{author}{\bibfnamefont{G.~F.} \bibnamefont{Franklin}},
  \bibinfo{author}{\bibfnamefont{J.~D.} \bibnamefont{Powell}},
  \bibnamefont{and}
  \bibinfo{author}{\bibfnamefont{A.}~\bibnamefont{Emami-Naeini}},
  \emph{\bibinfo{title}{Feedback control of dynamic systems}}, vol.
  \bibinfo{volume}{320} (\bibinfo{publisher}{Addison-Wesley},
  \bibinfo{address}{Reading, MA}, \bibinfo{year}{1991}).

\bibitem[{\citenamefont{{\O}ksendal}(2003)}]{oksendal2003stochastic}
\bibinfo{author}{\bibfnamefont{B.}~\bibnamefont{{\O}ksendal}},
  \emph{\bibinfo{title}{Stochastic differential equations}}
  (\bibinfo{publisher}{Springer Verlag}, \bibinfo{address}{Heidelberg,
  Germany}, \bibinfo{year}{2003}).

\bibitem[{\citenamefont{Vilar and Rub\'\i}(2001)}]{PhysRevLett.86.950}
\bibinfo{author}{\bibfnamefont{J.~M.~G.} \bibnamefont{Vilar}} \bibnamefont{and}
  \bibinfo{author}{\bibfnamefont{J.~M.} \bibnamefont{Rub\'\i}},
  \bibinfo{journal}{Phys. Rev. Lett.} \textbf{\bibinfo{volume}{86}},
  \bibinfo{pages}{950} (\bibinfo{year}{2001}).

\bibitem[{\citenamefont{Volpe et~al.}(2009)\citenamefont{Volpe, Wehr, Petrov,
  and Rubi}}]{volpe2009thermal}
\bibinfo{author}{\bibfnamefont{G.}~\bibnamefont{Volpe}},
  \bibinfo{author}{\bibfnamefont{J.}~\bibnamefont{Wehr}},
  \bibinfo{author}{\bibfnamefont{D.}~\bibnamefont{Petrov}}, \bibnamefont{and}
  \bibinfo{author}{\bibfnamefont{J.~M.} \bibnamefont{Rubi}},
  \bibinfo{journal}{J. Phys. A: Math. Theo.} \textbf{\bibinfo{volume}{42}},
  \bibinfo{pages}{095005} (\bibinfo{year}{2009}).

\bibitem[{\citenamefont{Kloeden and Platen}(1999)}]{kloeden}
\bibinfo{author}{\bibfnamefont{P.~E.} \bibnamefont{Kloeden}} \bibnamefont{and}
  \bibinfo{author}{\bibfnamefont{E.}~\bibnamefont{Platen}},
  \emph{\bibinfo{title}{Numerical solution of stochastic differential
  equations}} (\bibinfo{publisher}{Springer Verlag},
  \bibinfo{address}{Heidelberg, Germany}, \bibinfo{year}{1999}).

\bibitem[{\citenamefont{Volpe and Volpe}(2013)}]{volpe2013simulation}
\bibinfo{author}{\bibfnamefont{G.}~\bibnamefont{Volpe}} \bibnamefont{and}
  \bibinfo{author}{\bibfnamefont{G.}~\bibnamefont{Volpe}},
  \bibinfo{journal}{Am. J. Phys.} \textbf{\bibinfo{volume}{81}},
  \bibinfo{pages}{224} (\bibinfo{year}{2013}).

\bibitem[{\citenamefont{Ohta et~al.}(2007)\citenamefont{Ohta, Chiou, Phan,
  Sherwood, Yang, Lau, Hsu, Jamshidi, and Wu}}]{ohta2007optically}
\bibinfo{author}{\bibfnamefont{A.~T.} \bibnamefont{Ohta}},
  \bibinfo{author}{\bibfnamefont{P.-Y.} \bibnamefont{Chiou}},
  \bibinfo{author}{\bibfnamefont{H.~L.} \bibnamefont{Phan}},
  \bibinfo{author}{\bibfnamefont{S.~W.} \bibnamefont{Sherwood}},
  \bibinfo{author}{\bibfnamefont{J.~M.} \bibnamefont{Yang}},
  \bibinfo{author}{\bibfnamefont{A.~N.~K.} \bibnamefont{Lau}},
  \bibinfo{author}{\bibfnamefont{H.-Y.} \bibnamefont{Hsu}},
  \bibinfo{author}{\bibfnamefont{A.}~\bibnamefont{Jamshidi}}, \bibnamefont{and}
  \bibinfo{author}{\bibfnamefont{M.~C.} \bibnamefont{Wu}},
  \bibinfo{journal}{IEEE J. Sel. Top. Quant. El.}
  \textbf{\bibinfo{volume}{13}}, \bibinfo{pages}{235} (\bibinfo{year}{2007}).

\bibitem[{\citenamefont{Crocker and Grier}(1996)}]{crocker1996methods}
\bibinfo{author}{\bibfnamefont{J.~C.} \bibnamefont{Crocker}} \bibnamefont{and}
  \bibinfo{author}{\bibfnamefont{D.~G.} \bibnamefont{Grier}},
  \bibinfo{journal}{J. Colloid Interfac. Sci.} \textbf{\bibinfo{volume}{179}},
  \bibinfo{pages}{298} (\bibinfo{year}{1996}).

\bibitem[{\citenamefont{Rohrbach and Stelzer}(2002)}]{rohrbach2002three}
\bibinfo{author}{\bibfnamefont{A.}~\bibnamefont{Rohrbach}} \bibnamefont{and}
  \bibinfo{author}{\bibfnamefont{E.~H.~K.} \bibnamefont{Stelzer}},
  \bibinfo{journal}{J. Appl. Phys.} \textbf{\bibinfo{volume}{91}},
  \bibinfo{pages}{5474} (\bibinfo{year}{2002}).

\bibitem[{\citenamefont{Ebbens and Howse}(2010)}]{ebbens2010pursuit}
\bibinfo{author}{\bibfnamefont{S.~J.} \bibnamefont{Ebbens}} \bibnamefont{and}
  \bibinfo{author}{\bibfnamefont{J.~R.} \bibnamefont{Howse}},
  \bibinfo{journal}{Soft Matter} \textbf{\bibinfo{volume}{6}},
  \bibinfo{pages}{726} (\bibinfo{year}{2010}).

\bibitem[{\citenamefont{Bechinger et~al.}(2016)\citenamefont{Bechinger,
  Di~Leonardo, L{\"o}wen, Reichhardt, Volpe, and Volpe}}]{bechinger2016active}
\bibinfo{author}{\bibfnamefont{C.}~\bibnamefont{Bechinger}},
  \bibinfo{author}{\bibfnamefont{R.}~\bibnamefont{Di~Leonardo}},
  \bibinfo{author}{\bibfnamefont{H.}~\bibnamefont{L{\"o}wen}},
  \bibinfo{author}{\bibfnamefont{C.}~\bibnamefont{Reichhardt}},
  \bibinfo{author}{\bibfnamefont{G.}~\bibnamefont{Volpe}}, \bibnamefont{and}
  \bibinfo{author}{\bibfnamefont{G.}~\bibnamefont{Volpe}},
  \bibinfo{journal}{arXiv} p. \bibinfo{pages}{1602.00081}
  (\bibinfo{year}{2016}).

\bibitem[{\citenamefont{Berg and Brown}(1972)}]{berg1972chemotaxis}
\bibinfo{author}{\bibfnamefont{H.~C.} \bibnamefont{Berg}} \bibnamefont{and}
  \bibinfo{author}{\bibfnamefont{D.~A.} \bibnamefont{Brown}},
  \bibinfo{journal}{Nature} \textbf{\bibinfo{volume}{239}},
  \bibinfo{pages}{500} (\bibinfo{year}{1972}).

\bibitem[{\citenamefont{Mijalkov et~al.}(2016)\citenamefont{Mijalkov, McDaniel,
  Wehr, and Volpe}}]{PhysRevX.6.011008}
\bibinfo{author}{\bibfnamefont{M.}~\bibnamefont{Mijalkov}},
  \bibinfo{author}{\bibfnamefont{A.}~\bibnamefont{McDaniel}},
  \bibinfo{author}{\bibfnamefont{J.}~\bibnamefont{Wehr}}, \bibnamefont{and}
  \bibinfo{author}{\bibfnamefont{G.}~\bibnamefont{Volpe}},
  \bibinfo{journal}{Phys. Rev. X} \textbf{\bibinfo{volume}{6}},
  \bibinfo{pages}{011008} (\bibinfo{year}{2016}).

\bibitem[{\citenamefont{Volpe et~al.}(2014)\citenamefont{Volpe, Gigan, and
  Volpe}}]{volpe2014simulation}
\bibinfo{author}{\bibfnamefont{G.}~\bibnamefont{Volpe}},
  \bibinfo{author}{\bibfnamefont{S.}~\bibnamefont{Gigan}}, \bibnamefont{and}
  \bibinfo{author}{\bibfnamefont{G.}~\bibnamefont{Volpe}},
  \bibinfo{journal}{Am. J. Phys.} \textbf{\bibinfo{volume}{82}},
  \bibinfo{pages}{659} (\bibinfo{year}{2014}).

\end{thebibliography}

\end{document}